\documentclass[journal,onecolumn,12pt,twoside]{IEEEtran}
  \usepackage{setspace}
\usepackage{cite}
\usepackage{hyperref}

\ifCLASSINFOpdf
   \usepackage[pdftex]{graphicx}
\else
\fi

\usepackage{amsmath}
\usepackage{color}

\usepackage{url}

\begin{document}

\title{Present and Future of Reconfigurable Intelligent Surface-Empowered Communications}
%
%
%

\author{Ertugrul Basar,~\IEEEmembership{Senior Member,~IEEE} and H. Vincent Poor,~\IEEEmembership{Life Fellow,~IEEE}
\thanks{E. Basar is with CoreLab, Department of Electrical and Electronics Engineering, Ko\c{c} University, Istanbul, Turkey (e-mail: ebasar@ku.edu.tr).}
\thanks{H. Vincent Poor is with the Department of Electrical and Computer Engineering, Princeton University, Princeton, NJ 08544 USA (e-mail: poor@princeton.edu).} }

%
%

\markboth{IEEE Signal Processing Magazine}{Present and Future of Reconfigurable Intelligent Surface-Empowered Communications}
%



\maketitle


%

\vspace*{-1.5cm}
\section*{Introduction}
%
%
%
%
 \IEEEPARstart{S}{ignal} processing and communication communities have witnessed the rise of many exciting communication technologies in recent years. Notable examples include alternative waveforms, massive multiple-input multiple-output (MIMO) signaling, non-orthogonal multiple access (NOMA), joint communications and sensing, sparse vector coding, index modulation, and so on. It is inevitable that 6G wireless networks will require a rethinking of wireless communication systems and technologies, particularly at the physical layer (PHY), considering the fact that the cellular industry reached another important milestone with the development of 5G wireless networks with diverse applications \cite{Heath_2019}. Within this perspective, this article aims to shed light on the rising concept of \textit{reconfigurable intelligent surface} (RIS)-empowered communications towards 6G wireless networks \cite{Basar_Access_2019, Pan_2021}. Software-defined RISs can manipulate their impinging signals in an effective way to boost certain key performance indicators. We discuss the recent developments in the field and put forward promising candidates for future research and development. Specifically, we put our emphasis on active, transmitter-type, transmissive-reflective, and standalone RISs, by discussing their advantages and disadvantages compared to reflective RIS designs. Finally, we also envision an ultimate RIS architecture, which is able to adjust its operation modes dynamically, and introduce the new concept of PHY slicing over RISs towards 6G wireless networks.

\section*{Preliminaries}
An RIS is composed of many sub-wavelength and conductive elements, and it  can be viewed as a programmable cluster with many scatterers. As a whole, an RIS can be perceived as a signal processing system with multiple inputs and outputs, including several parallel subsystems with reconfigurable transfer functions.  

One of the fundamental advantages of RISs compared to existing solutions such as relaying and beamforming is their nearly passive nature. In other words, ideally, a passive RIS does not consume power for RF signal processing and signal manipulation. From a signals and systems perspective, the input-output relationship of the $n$th element of an RIS can be written as $y_n=p_n e^{j\phi_n}x_n$, where $x_n$ is the incoming signal, $y_n$ is the outgoing signal, and $\phi_n \in \left[0,2\pi \right) $ and $p_n \in \left[0,1 \right]  $ stand for the adjustable phase and amplitude of the $n$th RIS element, respectively. This phase adjustment can be implemented by means of variable loads, delay lines, or phase shifters. An RIS with $N$ elements can be thought of as $N$ parallel systems, whose adjustable parameters can be exploited for signal manipulation.

Signals incoming to an RIS are scattered to the propagation medium by the RIS elements, which results in a product path loss for the end-to-end link between the source and the destination. Specifically, the signals impinging on the RIS might suffer a considerable path loss and they further attenuate along their routes to their destination. Considering unit-gain (passive) RIS elements and line-of-sight (LOS) dominated links, one can easily show that the maximized received signal-to-noise ratio (SNR) for a passive RIS-empowered system follows \cite{SimRIS3}
\begin{equation}
\gamma_{\text{p}} \propto N^2\frac{P_T}{d_{SR}^{2} d_{RD}^{2} \sigma^2 }.
\end{equation}
Here, $P_T$ is the transmit power, $\sigma^2$ is the receiver noise power, and $d_{SR}$ and $d_{RD}$ respectively stand for source-RIS/RIS-destination distances, where the source-destination link is assumed to be unavailable. Two important conclusions can be drawn from (1). First, the received SNR increases with a factor of $N^2$, which has been known since the early RIS studies \cite{Basar_Access_2019}, thanks to the coherent combining of forward and backward links of the RISs with intelligent phase adjustment. However, the received SNR decays with a factor of $d_{SR}^{2} d_{RD}^{2}$, which considerably hurts the RIS-assisted systems in longer distances\footnote{For clarity of presentation, we dropped higher-order $\pi$ and wavelength $(\lambda$) terms in (1), while they cannot be ignored in a complete link-budget analysis.}. This is the reason why several research teams in our community, including ours, have been trying to find optimum RIS locations to boost their effect on the received signal. One conclusion of this work is that, if there exists a strong direct link, possibly with a LOS component that decays only with $d_{SD}^2$, where $d_{SD}$ is the source-destination distance, it becomes very challenging to find sweet spots to operate the RIS since its effect on the received SNR dwindles remarkably. In other words, considering the fact that the direct link SNR is given by $\gamma_{\text{d}} \propto P_T /(d_{SD}^2 \sigma^2)$, it would be challenging to compete with it if the RIS size is not extremely large and/or $d_{SR}$ and $d_{RD}$ are more than several tens of meters. Consequently, most recent practical campaigns generally position RISs either close to the transmitter or the receiver, to compensate this double path loss effect \cite{ScatterMIMO}. Unfortunately, we cannot foresee an easy solution to this low end-to-end path gain problem due to the underlying physics of RISs unless we consider active ones. 

In this article, we first introduce active RISs to overcome the double path loss problem of fully passive RISs. Then we put our emphasis on more sophisticated RIS architectures such as transmitter-type RISs, transmissive-reflective RISs, and standalone RIS, which might exploit the active RIS concept as well.

\section*{Passive vs. Active RISs}

The concept of active RISs has been recently introduced to overcome the aforementioned received power limitation of RISs \cite{Active_Passive_2021}. In this RIS architecture, RIS elements, equipped with active RF electronics, induce a power amplification factor $(p_n>1)$ for the impinging signals to boost the outgoing signals. From a signals and systems perspective, we have a new input-output relationship for this case: $y_n=p_n e^{j\phi_n}x_n + p_n e^{j\phi_n} v_n $, where $v_n \sim \mathcal{CN}(0,\sigma_v^2)$ is the input noise amplified by the $n$th active RIS element and thanks to the amplification, we have $p_n >1$. Here, $\mathcal{CN}$ stands for a circularly symmetrical complex Gaussian distribution.

 Assuming a maximum RIS-reflect power of $P_A$ and RIS-induced noise power of $\sigma_v^2$, the maximized received SNR can be obtained after simple derivations as

\begin{equation}
	\gamma_{\text{a}} \propto N\frac{P_T P_A}{P_A\sigma_v^2 d_{SR}^{2} + \sigma^2 d_{RD}^{2} + \sigma^2 \sigma_v^2 d_{SR}^{2} d_{RD}^{2}/N}.
\end{equation}
Here, thanks to an amplification that is dependent on source-RIS channels \cite{Active_Passive_2021}, one can escape from the fundamental limitation of RISs, multiplicative path-loss effect. This can be verified by a close inspection of (2) since the last term in the denominator can be readily dropped due to multiplicative noise variances. In other words, an active RIS can benefit from a superior path gain at the level of specular reflection that only experiences additive path loss.

In light of this information, in Fig. 1, we compare the received SNR of passive and active RIS-aided communication systems with the following parameters: $P_T=P_A=20$ dBm, $\sigma^2=\sigma_v^2=-90$ dBm, $N=256$, and for a carrier frequency of $5$ GHz. $xy$-coordinates of the source and RIS are respectively selected as $(0,0)$, and $(10,20)$, and the shown $12$ test points are given accordingly to assess the received SNR. For comparison purposes only, the received SNR of specular reflection $\gamma_{\text{s}} \propto P_T/((d_{SR}+d_{RD})^2 \sigma^2)$ is also calculated and shown in Fig. 1 without stressing too much on its practicality for the considered setup. We have the following observations: i) the received SNR for passive RIS diminishes very quickly when we get away from the RIS, ii) the active RIS not only provides less variation in received power but also significantly better (around $40$ dB higher) SNR thanks to an optimized amplification factor, which is calculated as $19.7$ dB for this specific setup\footnote{Following \cite{Active_Passive_2021}, we calculated the optimum amplification factor as $p_1=\cdots=p_N=\sqrt{\frac{P_A}{\sum_{n=1}^{N} \left| h_n\right|^2  + \sigma_v^2}}$. Here, $\left| h_n\right|$ is the magnitude of the $n$th source-RIS channel, which is equal to $\lambda/(4\pi d_{SR})$ for all $n$ under the far-field and LOS assumption.} and system parameters, and iii) in accordance with the amplification factor of the active RIS, it also provides approximately $20.6$ dB better SNR compared to the specular reflection\footnote{This can be also verified by the ratio $\gamma_{\text{a}}/\gamma_{\text{s}} 	\simeq \frac{NP_A(d_{SR}+d_{RD})^2}{(P_A d_{SR}^2 + d_{RD}^2)}\sim N$ assuming $\sigma_v^2=\sigma^2 $ and $N\gg 1$.}. 

\begin{figure}[t]
	\centering
	\includegraphics{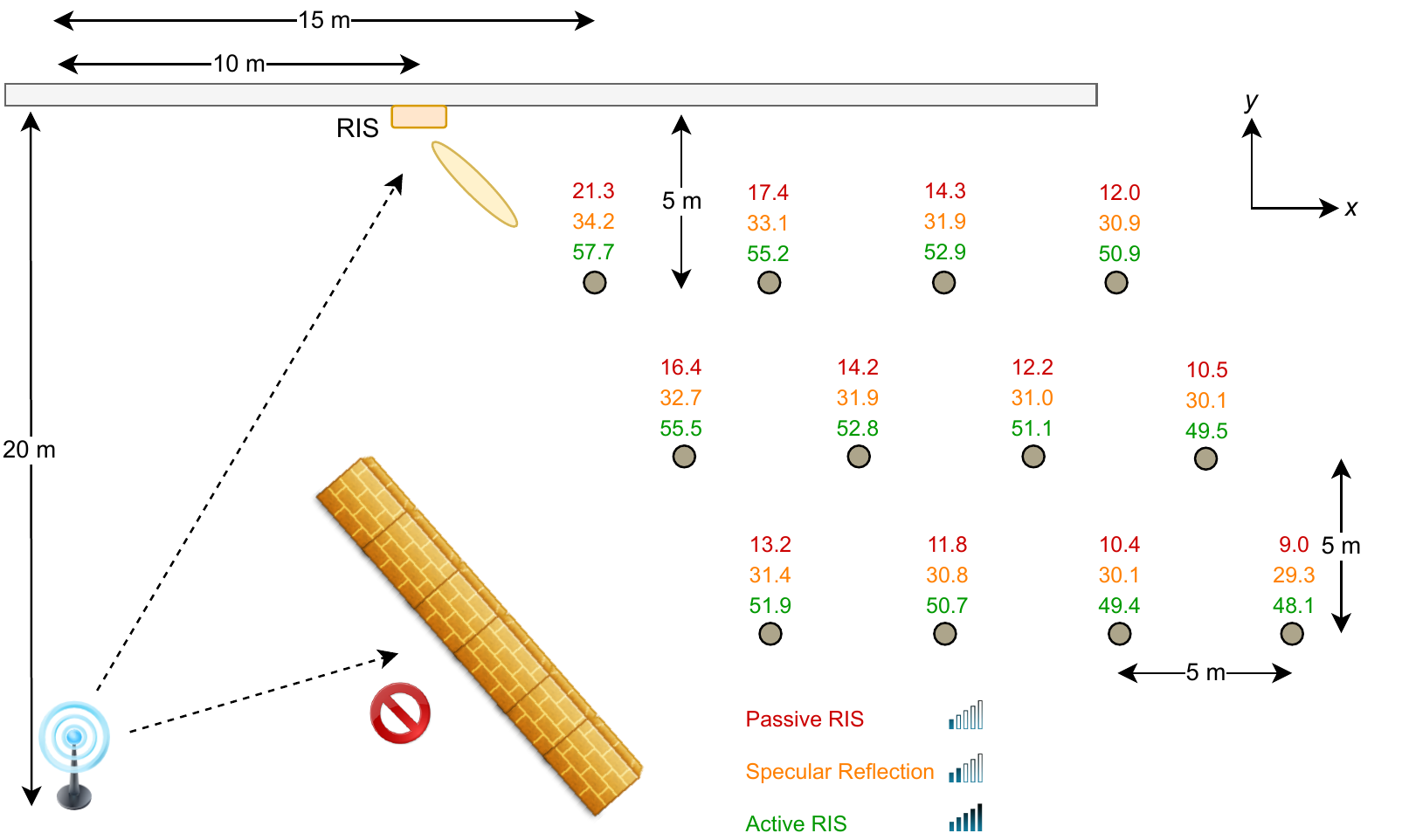}
	\caption{Comparison of passive RIS, specular reflection, and active RIS in terms of received SNR in decibels.}
\end{figure}

In conclusion, the active RIS not only transforms the multiplicative path loss of passive RISs into an additive form, but also benefits from an additional amplification gain on top it, making active RIS systems even more more powerful than specular reflection alone. A hybrid architecture comprising both active and passive RIS elements might further improve the analysis and optimization, and can provide a compromise between two solutions. The power consumption associated with active RIS elements might be a concern for RISs not connected to the power grid, and a detailed investigation of these issues is beyond the scope of this article. We envision that an active RIS might outperform an amplify-and-forward relay in terms of cost and efficiency, and leave this interesting comparison to future studies.

\section*{Relay- vs. Transmitter-Type RISs}
The conceptual similarity between relaying and RIS-assisted systems has attracted the attention of many researchers recently. While we believe that an RIS provides many distinguishing  features compared to a half- or full-duplex relay, depending on its functionality, it can operate in either relay or transmitter mode. As shown in Fig. 1, in relay mode, the RIS (whether active or passive) is placed somewhere between the source and the destination to assist the communication in between, most probably, when the direct link is very weak or blocked. However, if we decide to follow the path of passive RISs, a very promising direction for RISs might be transmitter-type operation, where the RIS becomes a part of the transmitter and plays an active role in signaling and modulation \cite{Basar_2019_LIS}.

In a setup where the RIS is used in the transmitter mode, it can manipulate the incoming unmodulated carrier signal to encode information bits. In other words, by feeding the RIS with an unmodulated carrier, it is possible to create virtual signal constellations over-the-air to convey information. This can be verified from the RIS element's basic signal model $y_n=p_n e^{j\phi_n} x_n$, where with proper adjustments of $p_n$ and $\phi_n$, virtual amplitude-phase modulation constellations can be created. Following the popular terminology of \textit{modulations}, notable examples include spatial modulation, media-based modulation, and the more general index modulation, transmitter-type RISs can be regarded as reflection modulation, where we embed information into reflection states of an RIS. In principle, this is conceptually a similar architecture to media-based modulation \cite{Basar_2019}, where different radiation patterns of a reconfigurable antenna is exploited to transmit information, that is, to create a virtual signal constellation. 

Two major advantages of transmitter-type RISs are summarized as follows. First, the hardware architecture of an RIS-based transmitter would be much simpler compared to a traditional one with upconverters and filters, while an unmodulated cosine carrier signal can be generated very easily using an RF digital-to-analog converter with an internal memory and a power amplifier. Second, since the RIS is relatively close to the signal source, the multiplicative path loss effect is restrained. One can even position the RIS to lie in the near-field of the transmitter to further boost the received signal power, however, taking into account near-field effects carefully.

To sum up, transmitter-type RISs might be a remedy to implement RF chain-free transmitters and to mimic MIMO systems by transmitting multiple streams simultaneously over the RIS with a relatively simple architecture.  We refer interested readers to the recent study of \cite{Jin_2021}, and the references therein, for practical transmitter-type RIS architectures developed by researchers from Southeast University.

\section*{Reflective Only vs. Transmissive-Reflective RISs}
Is it possible to provide coverage on the back side of an RIS? The answer is yes, thanks to the simultaneously transmitting and reflecting RISs \cite{Star_2021,Star_2021_2}. Our discussion so far has generally assumed that both the source and the destination are on the same side of the RIS, that is, within the same half space of the smart environment. This is a constraint stemming from the physical architecture of the RIS, which is composed of a conductive substrate that acts as a blocker for the signals hitting the surface, not allowing them to pass through the surface. The concept of transmissive-reflective RISs challenges this status quo by carefully reengineering the RIS structure to allow it to simultaneously reflect and transmit (in principle, refract) the incoming signals to provide a full $360^{\text{o}}$ coverage. This is achieved by magnetic RIS elements that can support not only surface
electric polarization currents but also magnetic currents to simultaneously control the refracted and reflected signals.

Early research on RISs envisioned them as objects hanging on walls or facades of buildings. But, an RIS could also be positioned in the middle of a communication environment, or embedded in a wall between two different environments, to receive signals and forward them in different directions. From a signal processing perspective, this type of RIS can be considered as a single-input dual-output system, however, a number of problems need to be solved in terms of system protocol design, such as mode, energy, and time splitting. Other concerns include the complicated hardware architecture and potentially unaesthetic appearance of such RISs, which might be the price to be paid for a full $360^{\text{o}}$ coverage. Nevertheless, such surfaces can provide effective solutions for outdoor-to-indoor coverage extension, multi-room coverage, and realizing more effective NOMA 2.0 systems. It is worth noting that the multiplicative path loss effect discussed earlier would still be a concern with this type of design and might potentially limit the overall coverage on both sides of the RIS. In this context, the design of an active transmissive-reflective surface would be a notable leap forward for full-coverage.

We conclude this section by quoting the authors of \cite{Star_2021} for the comparison of reflective only and transmissive-reflective RISs: biscuits placed on a metal plate and ice cubes in a glass of water, which correspond to reflective RIS elements on an opaque RIS substrate and magnetic RIS elements on a transparent RIS substrate, respectively.  Which will prevail in the following years? The answer will be given in the following years considering the interesting trade-offs that they offer in terms of coverage and complexity in smart environments.

\section*{Interconnected vs. Standalone RISs}
Some key questions that arise is research on RISs include the following:  How might an RIS obtain knowledge of channel phases? How can the RIS adjust itself in real-time? Is it possible to perform channel estimation at the RIS? How realistic is it to assume fully-passive RISs? Where does the intelligence of the RIS come from? We have to admit that these questions are not easy to answer, but thanks to recent developments in the field, a number of interesting directions might be followed to realize smart RISs for 6G wireless networks.

A fully-passive but interconnected RIS has a dedicated communication link, that is, a fully-functional RF chain, that receives critical information to be forwarded to the microprocessor controlling the RIS. This link might be between the RIS and the source, destination, or a central control unit and acts as a guide for the RIS when adjusting its reflection state. On the other hand, a standalone RIS is equipped with sparse sensors embedded among passive RIS elements and has a number of RF chains for background baseband signal processing to acquire knowledge about the wireless environment. A notable example is given in \cite{Taha_2021}, where the use of sparse sensors is introduced in the first place by considering a hybrid RIS architecture with a number of elements connected to the baseband. The use of deep learning tools at the RIS to train it on how to interact with incoming signals is another very promising direction for standalone RISs. Here, the considered deep learning model learns how to map
the observed environment descriptors (channels) to the optimal RIS configurations. 

We finally note that a detailed investigation of potential artificial intelligence techniques for RIS-empowered systems might be the topic of another article. It would not be surprising to see that future RISs will leverage tools from artificial intelligence to adapt themselves in real time without any user intervention.

\section*{The Ultimate RIS Architecture and PHY Slicing}
Our discussion so far has provided interesting perspectives by focusing on a  number of promising RIS architectures and making relative comparisons among them. To provide the bigger picture with further insights, the considered RIS architectures in this article are summarized in Fig. 2 with the pros and cons of each of them, where we focus on passive vs. active RISs, relay- vs. transmitter-type RISs, reflective-only vs. transmissive-reflective RISs, and interconnected vs. standalone RISs, respectively. We believe that the future research and development in this field will possibly focus on these interesting RIS architectures, particularly the ones at the right-hand side of Fig. 2, trying to unlock their true potential by exploring new applications and low-cost designs. We also note that all four architectures at the left-hand side of Fig. 2 are envisioned to have a passive nature initially, while it is also possible to use the active RIS concept for transmitter-type and standalone RISs, and potentially transmissive-reflective RISs by carefully reengineering RIS elements. 

\begin{figure}[!t]
	\centering
	\includegraphics[width=0.85\columnwidth]{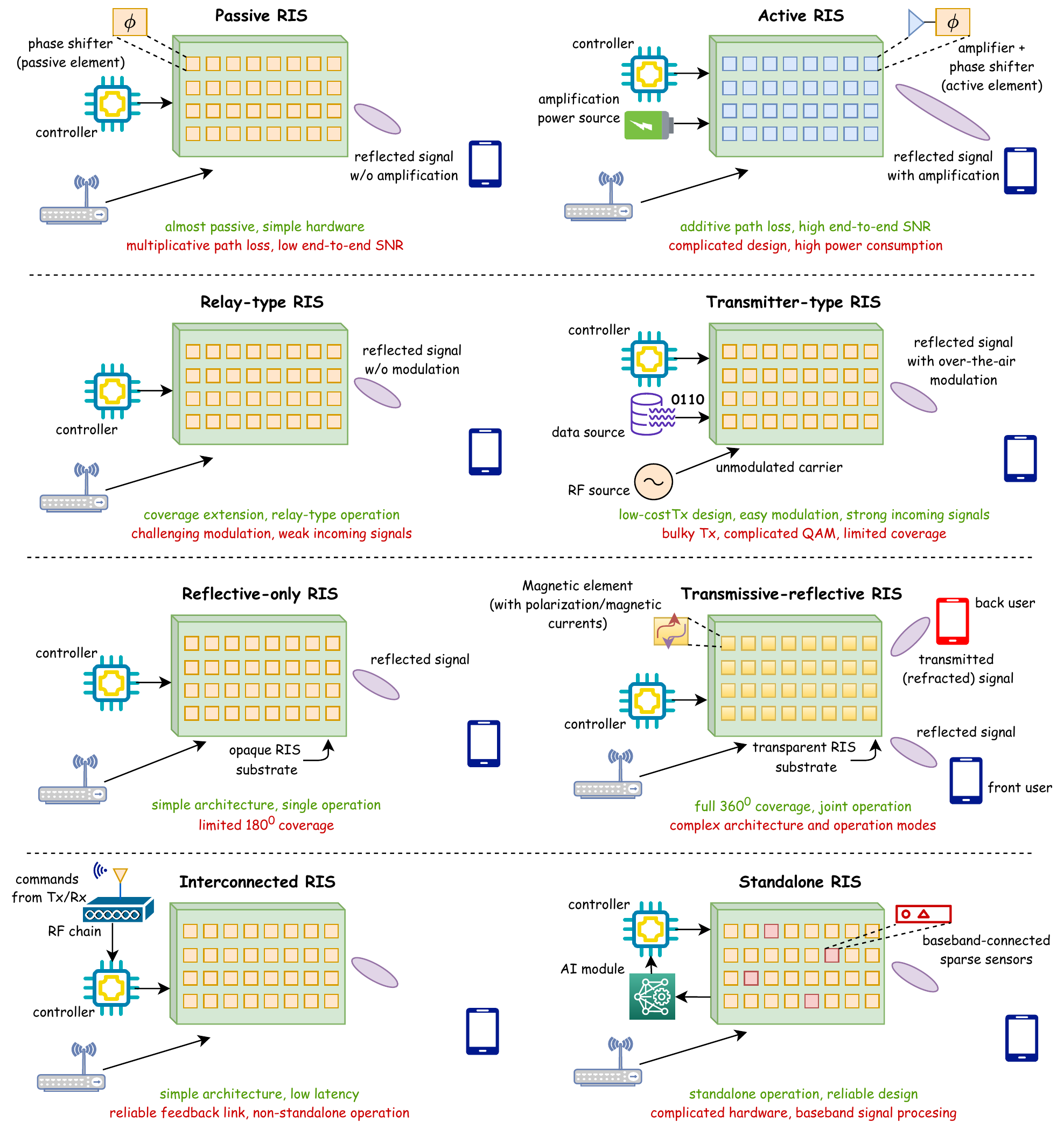}
	\caption{Comparison of different RIS architectures with their pros and cons.}
\end{figure}

Against this background, we propose an ultimate RIS architecture shown in Fig. 3. In this RIS design, we envision a combination of the aforementioned RIS architectures while assuming that these type of sophisticated RISs might be constructed physically in the future. Here, the ultimate RIS has a hybrid architecture with a mix of passive and active RIS elements, embedded with sparse sensors for channel acquisition, and has a dedicated subsurface for transmissive-reflective operation. To eliminate interference and/or improve secrecy, this RIS might also switch to the absorption mode without reflecting or transmitting any signal. In particular, the ultimate RIS can simultaneously work in the following modes: active-reflect, active-transmit, passive-reflect, passive-transmit, and absorb. The ultimate RIS might also collect sensory information from its neighborhood and convey this data with transmitter-type operation (modulation on-the-fly) if necessary. Advanced functionalities of this ultimate RIS might include over-the-air Doppler/multipath mitigation, channel equalization, and interference cancellation. Although the ultimate RIS architecture might be perceived as a straightforward combination of existing RIS technologies, the underlying signal processing tasks and communication engineering problems would necessitate novel approaches. For instance, an interesting optimization problem needs to be solved to determine the optimal sizes of  different subsurface slices. Furthermore, this RIS should be carefully positioned considering the distribution of near, far, and back users to achieve the maximum sum-rate.

\begin{figure}[!t]
	\centering
	\includegraphics[width=0.7\columnwidth]{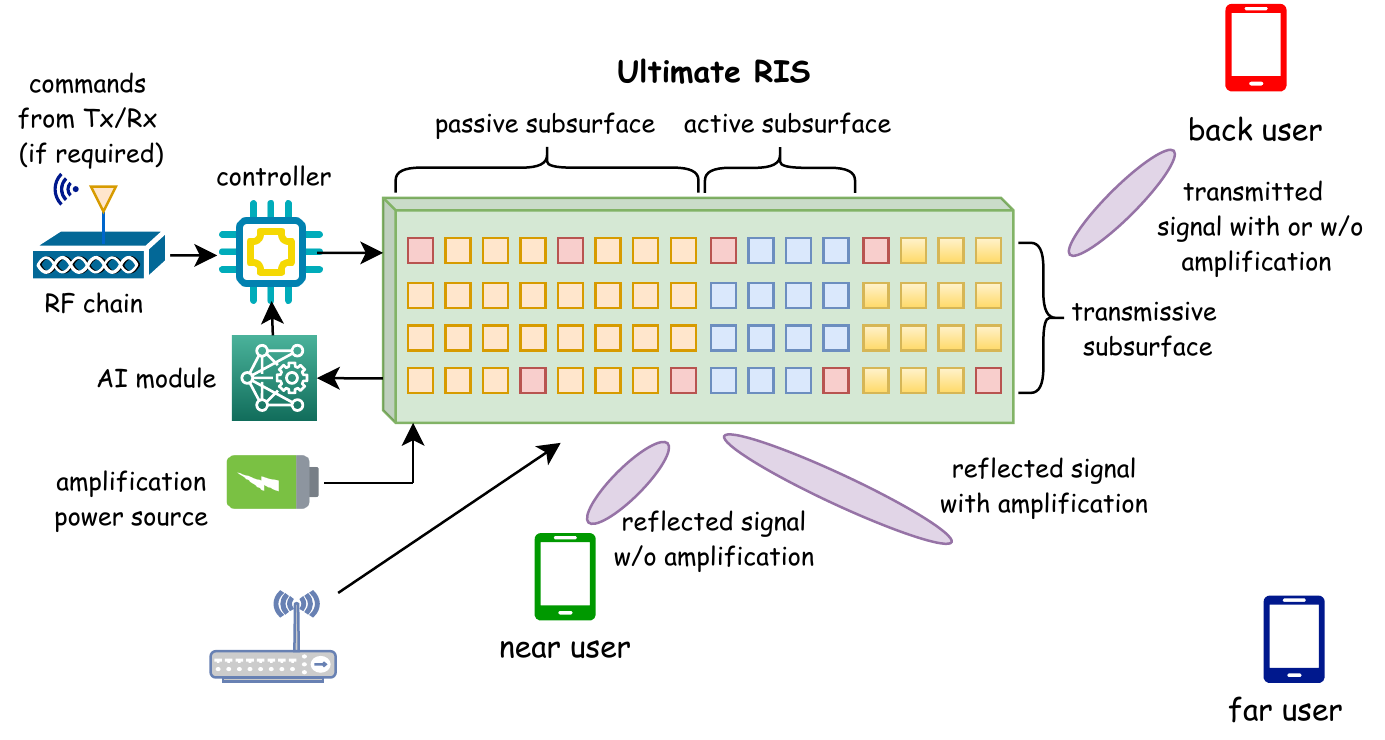}
	\caption{The ultimate RIS architecture composed of different subsurfaces for improved flexibility.}
\end{figure}

The ultimate RIS architecture shown in Fig. 3 might be exciting in its own right, but it may further pave for a new concept towards 6G: \textit{PHY slicing}. Inspired by the 5G network slicing concept, RIS-empowered PHY slicing would enable the multiplexing of different RIS architectures and/or applications on the same physical RIS by dividing the RIS into subsurfaces. This RIS with PHY slicing functionality might be fully reconfigurable, meaning that the sizes of passive, active, and transmissive subsurfaces could be adjusted in real-time according to the needs of the users in the network. Different applications, such as data transfer, wireless charging, PHY security etc., could be simultaneously supported by this ultimate RIS as well by slicing the RIS.

\section*{Concluding Remarks}
In this article, we have provided a general perspective on the future of RIS technologies towards 6G wireless networks and put forward an ultimate RIS architecture. We have also shown that an active RIS might be a feasible solution to the low path gain problem of passive RISs by a quantitative analysis in terms of received SNR. At this point, we can raise several questions for future research: Can we ignore the power consumption of active RISs, and more importantly, might the increasing electromagnetic exposure stemming from the active components be a concern? Will an active RIS be more cost effective than relays or small base stations? Who will control and monitor RISs   (users, vendors, operators?), particularly the standalone ones? Do we need a single and sophisticated RIS or simpler and smaller ones employed in large numbers? How many RISs are required then to cover a certain area? We hope to find answers to these interesting questions in the near future. 

We conclude that passive RISs might provide cost-effective solutions to increase indoor coverage and overcome blockage problems, while active ones would be more suitable for outdoor environments in which distances between terminals are much larger and RIS power consumption might not be a major concern. Transmitter-type RISs might be a remedy for future Internet-of-things applications with their relatively simple transmitter architectures without any RF chains. Finally, transmissive-reflective RISs may unlock the potential of future NOMA systems by also providing outdoor-to-indoor coverage extension.

Clearly, the concept of RIS-empowered communication has brought a great deal of excitement to our community recently and we will see whether it will be a strong candidate towards 6G in the following years.

\section*{Open Codes}

The MATLAB script used to obtain the SNR values in Fig. 1 is given at \url{https://corelab.ku.edu.tr/tools/}.

\section*{Acknowledgment}

The work of E. Basar was supported by the Scientific and Technological Research Council of Turkey (TUBITAK)-COST project 120E401.

The work of H. V. Poor was supported by the U.S. National Science Foundation under Grant CCD-1908308.

\ifCLASSOPTIONcaptionsoff
  \newpage
\fi



%

\bibliographystyle{IEEEtran}
\bibliography{bib_2020}

%

\newpage

\begin{IEEEbiographynophoto}{Ertugrul Basar} received his Ph.D. degree from Istanbul Technical University in 2013. He is currently an Associate Professor with the Department of Electrical and Electronics Engineering, Ko\c{c} University, Istanbul, Turkey and the director of Communications Research and Innovation Laboratory (CoreLab). His primary research interests include beyond 5G systems, index modulation, intelligent surfaces, waveform design, and signal processing for communications. Dr. Basar currently serves as a Senior Editor of \textit{IEEE Communications Letters} and an Editor of \textit{IEEE Transactions on Communications} and \textit{Frontiers in Communications and Networks}. He is a Young Member of Turkish Academy of Sciences and a Senior Member of IEEE.
\end{IEEEbiographynophoto}

\vspace*{-18cm}

\begin{IEEEbiographynophoto}{H. Vincent Poor} is the Michael Henry Strater University Professor at Princeton University. His interests include information theory, machine learning and networks science, and their applications in wireless networks, energy systems, and related fields. Dr. Poor is a member of the National Academy of Engineering and the National Academy of Sciences, and a foreign member of the Chinese Academy of Sciences and the Royal Society. He received the Technical Achievement and Society Awards of the IEEE Signal Processing Society in 2007 and 2011, respectively, and the IEEE Alexander Graham Bell Medal in 2017.
\end{IEEEbiographynophoto}





\end{document}